\documentclass[onecolumn,english,pra,showpacs,superscriptaddress,longbibliography]{revtex4-1}
\usepackage{amsfonts}
\usepackage{amsmath}
\usepackage{amssymb}
\usepackage{amsthm}
\usepackage[colorlinks, citecolor=blue,linkcolor=red]{hyperref}
\usepackage{color}
\usepackage{hyperref}
\usepackage{textcomp}
\usepackage[rightcaption]{sidecap}
\usepackage{subfigure}
\usepackage[rightcaption]{sidecap}
\usepackage{graphicx}
\usepackage[utf8]{inputenc}
\usepackage[T1]{fontenc}
\usepackage{lipsum}
\begin{document}
\title{Spectroscopic readout of chiral photonic topology in a single-cavity spin-orbit-coupled Bose–Einstein condensate}
\author{Kashif Ammar Yasir}
\email {kayasir@zjnu.edu.cn}\affiliation{Department of Physics, Zhejiang Normal University, Jinhua 321004, China.}
	\affiliation{Zhejiang Institute of Photoelectronics, Jinhua 321004, China.}
\author{Wu-Ming Liu}
\email{wliu@iphy.ac.cn}
\affiliation{Beijing National Laboratory for Condensed Matter Physics, Institute of Physics, Chinese Academy of Sciences, Beijing 100190, China.}
\author{Gao Xianlong}
\email {gaoxl@zjnu.edu.cn}\affiliation{Department of Physics, Zhejiang Normal University, Jinhua 321004, China.}

\setlength{\parskip}{0pt}
\setlength{\belowcaptionskip}{-10pt}
\begin{abstract}
	Topological photonic phases are typically identified through band reconstruction, steady-state transmission, or real-space imaging of edge modes. In this work, we present a framework for spectroscopic readout of chiral photonic topology in a single driven optical cavity containing a spin-orbit-coupled Bose-Einstein condensate. We demonstrate that the cavity transmission power spectral density provides a direct and measurable proxy for a momentum- and frequency-resolved photonic Chern marker, enabling topological characteristics to be inferred from spectral data without the need for bulk-band tomography. In the loss-dominated regime, where cavity decay exceeds atomic dissipation, the power spectral density exhibits Dirac-like gapped hybrid modes with a vanishing Chern marker, indicating a trivial phase. When the dissipation imbalance is reversed, a bright, gap-spanning spectral ridge emerges, co-localized with peaks in both the Chern marker and Berry curvature. The complex spectrum reveals parity-time symmetric coalescences and gain-loss bifurcations, marking exceptional points and enabling chiral, gap-traversing transport. By linking noise spectroscopy to geometric and non-Hermitian topology in a minimal cavity-QED architecture, this work provides a framework for spectroscopic detection of topological order in driven quantum systems. This approach offers a pathway to compact, tunable topological photonics across a broad range of light-matter platforms, providing a method for the study and control of topological phases in hybrid quantum systems.
\end{abstract}
\date{\today}
\maketitle

Topological photonics has enabled unprecedented control of light by exploiting global geometric phases rather than local material properties~\cite{Lu2014,Ozawa2019}. Analogues of quantum Hall and quantum spin Hall insulators have been realized in photonic crystals, coupled resonator arrays and metamaterials, revealing robust chiral edge modes that propagate without backscattering~\cite{Hafezi2013,Rechtsman2013,Khanikaev2013,Barik2018}. More recently, attention has shifted to non-Hermitian and driven photonic systems, where balanced gain and loss, parity--time (PT) symmetry and exceptional points (EPs) give rise to topological phases without Hermitian counterparts~\cite{ElGanainy2018,Ozdemir2019,Cerjan2019,Liu2025}. In these platforms, the real and imaginary parts of complex eigenvalues jointly determine topology, yet most experimental probes rely on steady-state transmission or band tomography and provide limited access to \emph{local} signatures such as Berry curvature or real-space Chern markers~\cite{Parto2021,Hu2021,Miri2019}. A method to extract topological information directly from cavity output fields, including fluctuation and noise spectra, is still lacking.

Ultracold atoms in optical cavities provide a natural interface between quantum optics and topological matter~\cite{Ritsch2013,Mivehvar2021}. In particular, spin--orbit coupled (SOC) Bose--Einstein condensates (BECs) mimic solid-state topological phases, exhibiting Dirac dispersions, magnetic textures and Berry curvature hotspots~\cite{Lin2011,Galitski2013,Zhai2015,Lian2017}. When placed inside a high-finesse cavity, such atoms hybridize with the cavity mode, allowing their internal pseudo-spin structure to be imprinted onto light~\cite{Brennecke2007,Padhi2014,Deng2014,Ostermann2021}. This has enabled observations of cavity-induced magnetism, superradiant phase transitions and optomechanical backaction modified by SOC~\cite{Hamner2014,Yasir2017,Landig2015,yasir2023,yasir2024}. Topological features have been inferred from mean-field cavity transmission, electromagnetically induced transparency (EIT) and Raman spectroscopy~\cite{Clerk2010,Jia2023,Yasir2022,Qin2021,Peng2014}. However, these approaches primarily access average transmission or global Chern numbers in quasi-Hermitian regimes, and do not resolve how topology, dissipation and quantum fluctuations manifest in the \emph{power spectral density} (PSD) of the cavity output. Moreover, the role of non-Hermitian band topology, Berry curvature and exceptional points in cavity-coupled SOC BECs remains largely unexplored~\cite{Minganti2019,Lieu2018,Chen2022}.

In this article, we demonstrate that cavity transmission PSD offers a direct, experimentally accessible probe of non-Hermitian topology in a driven SOC BEC–cavity system. By computing the momentum- and frequency-resolved PSD of the output field, including quantum and thermal fluctuations, we construct a photonic Chern marker, a local analogue of the Chern number that characterizes topology without requiring integration over the full Brillouin zone \cite{Bianco2011, Huang2014, Marzari2012}. When cavity decay dominates atomic dissipation ($\kappa > \gamma$), the hybrid atom-photon spectrum exhibits gapped Dirac-like bands, and the Chern marker forms localized hotspots near avoided crossings. In contrast, when $\gamma > \kappa$, the system enters a $\mathcal{PT}$-symmetric gain-loss regime, where real and imaginary eigenvalues coalesce at exceptional points and then bifurcate to form ring-shaped gapless contours around the pseudo-spin manifolds \cite{ElGanainy2018, Cerjan2019}. The corresponding Chern marker reorganizes into annular, sign-alternating structures, tracing edge-like photonic modes. Additionally, we reconstruct the Berry curvature from the complex band structure inferred from the PSD and show its correspondence with the Chern marker distribution. Our work establishes cavity transmission noise as a powerful tool for observing Chern markers, Berry curvature, and exceptional-point physics in hybrid atom-photon systems, offering a pathway toward topological sensing and information processing with quantum gases in cavities.

\section*{Results}
\subsection*{System Description}
We analyze a spin–orbit–coupled Bose–Einstein condensate embedded in a single–mode optical resonator, motivated by the $^{87}\mathrm{Rb}$ implementations of Ref.~\cite{Lin2011}. The atomic medium is a BEC with $N \simeq 1.8 \times 10^{5}$ atoms confined inside a high–$Q$ Fabry–Pérot cavity of length $L \simeq 12.5 \times 10^{-3}~\mathrm{m}$, see Fig.~\ref{Fig1}(a). The cavity axis is taken along $\hat{x}$ and supports a single mode of frequency $\omega_c \simeq 1.9 \times 2\pi~\mathrm{GHz}$ with photon loss rate $\kappa$. A laser of power $P$ drives this mode along $\hat{x}$ with amplitude $\vert \eta \vert = \sqrt{P\kappa/(\hbar\omega_E)}$, where the drive frequency is written as $\omega_E = \omega_R + \delta\omega_R$ to make explicit its relation to the Raman fields. The corresponding detuning between the pump and the bare cavity resonance is $\Delta_c = \omega_E - \omega_c$, and we operate in a near–dispersive regime with $\Delta_c \approx \kappa$, such that $\omega_c \simeq \omega_R + \delta\omega_R$ in the limit $\kappa \rightarrow 0$. Such a hybrid cavity--SOC configuration is experimentally plausible at the effective-model level, since cavity-coupled quantum gases and cavity-assisted spin--orbit coupling have already been realized in closely related cold-atom platforms~\cite{Landig2015,Leonard2017,Kroeze2018,Norcia2018,Kroeze2019}. In this sense, the present geometry should be viewed as an experimentally motivated effective description in which the cavity field provides the dynamical lattice, while the Raman sector supplies the SOC coupling.

A homogeneous magnetic bias field of magnitude $B_0 \simeq 10~\mathrm{G}$ is applied perpendicular to the cavity axis (in the $y$--$z$ plane), producing a Zeeman splitting $\hbar \omega_z$ between selected hyperfine levels and ensuring $\lvert \omega_z/\kappa \rvert \gg 1$. The pseudo–spin degree of freedom is formed by the two internal states $\vert \uparrow \rangle = \vert F = 2, m_F = 0 \rangle$ and $\vert \downarrow \rangle = \vert F = 2, m_F = -1 \rangle$ within the $F = 2$ manifold. SOC is realized using two counter–propagating Raman beams of frequencies $\omega_R$ and $\omega_R + \delta\omega_R$ and wavelength $\lambda = 804.1~\mathrm{nm}$, propagating along $\hat{x}$ with wave vectors $\pm k_x \hat{x}$, where $k_x = 2\pi/\lambda$. The single–photon recoil energy is $E_x = \hbar^2 k_x^2/(2m_a)$ for atomic mass $m_a$, and the two–photon Raman detuning is chosen as $\delta \simeq 1.6 E_x$, in accordance with Ref.~\cite{Lin2011}. In this geometry the Raman process ties the atomic spin to the center–of–mass motion along $\hat{x}$, generating an equal Rashba–Dresselhaus SOC in which the two spin components experience shifted dispersion minima and momentum–dependent spin mixing.

The single–particle dynamics of the pseudo–spin states are captured by the SOC Hamiltonian
$
\hat{\mathcal{H}}_{SOC} 
= \frac{\hbar^2 q_x^2}{2m_a} \sigma_{0}
+ \tilde{\alpha} q_x \sigma_{y}
+ \frac{\delta}{2} \sigma_y
+ \frac{\Omega_z}{2} \sigma_z ,
$
where $q_x$ denotes the atomic quasimomentum entering the effective SOC Hamiltonian, while $k_x=2\pi/\lambda$ remains the fixed recoil wave vector set by the Raman wavelength. Here $\sigma_0$ is the $2 \times 2$ identity, $\sigma_{x,y,z}$ are Pauli matrices acting in the $\{\vert \uparrow \rangle, \vert \downarrow \rangle\}$ basis, and the SOC strength is $\tilde{\alpha} = E_x/k_x = \hbar^2 k_x/(2m_a)$. The parameters $\delta$ and $\Omega_z$ are expressed as $\delta = - g \mu_B B_z$ and $\Omega_z = - g \mu_B B_y$, with $g$ the Landé $g$–factor, $\mu_B$ the Bohr magneton, and $B_y$ and $B_z$ the components of the bias field along $\hat{y}$ and $\hat{z}$, respectively. In this language, $\delta$ encodes a tunable Zeeman field along $\hat{y}$, while $\Omega_z$ represents an effective Zeeman field along $\hat{z}$ controlled by the bias–field orientation.

Within the rotating–wave approximation, the many–body dynamics of the coupled BEC–cavity system are described by
\begin{eqnarray}\label{Ha1}
	\hat{\mathcal{H}} &=& 
	\int dx\,\pmb{\hat{\psi}}^{\dag}(x)
	\big[\hat{\mathcal{H}}_{0}+\mathcal{V}(x)\big]
	\pmb{\hat{\psi}}(x)
	+ \frac{1}{2} \int dx \sum_{\sigma,\sigma'} 
	\mathcal{U}_{\sigma\sigma'}\,
	\hat{\psi}^{\dag}_\sigma
	\hat{\psi}^{\dag}_{\sigma'}
	\hat{\psi}_{\sigma'}
	\hat{\psi}_\sigma \nonumber \\[2pt]
	&& + \hbar \Delta_c\, \hat{c}^{\dag} \hat{c}
	- i \hbar \eta (\hat{c} - \hat{c}^{\dag}),
\end{eqnarray}
where $\pmb{\hat{\psi}} = (\hat{\psi}_{\uparrow}, \hat{\psi}_{\downarrow})^T$ denotes the two–component bosonic field operator, $\hat{c}$ ($\hat{c}^{\dag}$) annihilates (creates) a cavity photon, and $\hat{\mathcal{H}}_{0} \equiv \hat{\mathcal{H}}_{SOC}$ is the single–particle Hamiltonian defined above. 

The atoms are subject to a dispersive cavity-induced optical potential $\mathcal{V}(x)=\hbar U_0\,\hat{c}^{\dag}\hat{c}\cos^2(k_c x)$, where $U_0=g_0^2/\Delta_a$ is the light shift per intracavity photon (with $g_0$ the single--photon Rabi frequency and $\Delta_a$ the atom--cavity detuning) and $k_c$ denotes the cavity wave vector. This potential originates from the standing-wave cavity mode and is dynamically controlled by the intracavity photon number $\hat{c}^{\dag}\hat{c}$. In the collinear configuration considered here, the cavity field and the Raman beams propagate along the same axis but are spectrally and polarization selective: the control field populates the cavity mode, while the Raman fields implement the two-photon coupling between $\vert\uparrow\rangle$ and $\vert\downarrow\rangle$ with negligible cavity excitation. Any scalar AC Stark shift from the Raman beams is independent of $\hat{c}^{\dag}\hat{c}$ and can be absorbed into the chemical potential (or treated as a weak static background), so $\mathcal{V}(x)$ provides the relevant dynamical lattice potential that enters our effective description~\cite{Lin2011,Brennecke2007,haldane1988,Kane2005,Hasan2010}.

Contact interactions between atoms in spin components $\sigma,\sigma' \in \{\uparrow,\downarrow\}$ are modeled by 
$\mathcal{U}_{\sigma \sigma'} = 4\pi \hbar^2 a_{\sigma \sigma'}/m_a$,
where $a_{\sigma \sigma'}$ denote the $s$–wave scattering lengths. This form incorporates both density–density and spin–exchange channels within the SOC BEC coupled to the dynamical cavity field.

We assume $U_{\uparrow\uparrow} = U_{\downarrow\downarrow} = U$ and 
$U_{\uparrow\downarrow} = U_{\downarrow\uparrow} = \varepsilon U$. 
Using the plane-wave ansatz 
$\pmb{\hat{\psi}}(x) = e^{i\pmb{k} \cdot x} \pmb{\hat{\varphi}}$, 
with $\pmb{\hat{\varphi}} = (\hat{\varphi}_{\uparrow}, \hat{\varphi}_{\downarrow})^T$ normalized as 
$|\hat{\varphi}_{\uparrow}|^2 + |\hat{\varphi}_{\downarrow}|^2 = N$, 
the dynamics reduce to the coupled quantum–Langevin equations,
\begin{eqnarray}\label{QLE}
	\dot{\hat c}&=&(i\Delta_a-ig_{a}\pmb{\hat{\varphi}}^{\dag}\pmb{\hat{\varphi}}-\kappa)\hat{c}
	+\eta+\sqrt{2\kappa}\,a_{\mathrm{in}},\nonumber\\
	\dot{\pmb{\hat{\varphi}}}
	&=&\Big[\frac{\hbar \pmb{k}^2}{2m_a}\sigma_{0}
	+\tilde{\alpha}k_x\sigma_y
	+\frac{\delta}{2}\sigma_y
	+\frac{\Omega_z}{2}\sigma_z
	-\gamma
	+g_a\hat{c}^{\dag}\hat{c}\Big]\pmb{\hat{\varphi}}\nonumber\\
	&&+\frac{U}{2}\pmb{\hat{\varphi}}^{\dag}\pmb{\hat{\varphi}}\,\pmb{\hat{\varphi}}
	+\frac{\varepsilon U}{2}\hat{\varphi}_{\sigma}^{\dag}\hat{\varphi}_{\sigma'}\hat{\varphi}_{\sigma}
	+\sqrt{2\gamma}\,f_a,
\end{eqnarray}
where $a_{\mathrm{in}}$ and $f_a$ are the input noise operators of the cavity and atomic fields. In the discussion below, the momentum label $k_x$ is retained as the spectroscopic coordinate used in the cavity-transmission and edge-mode analysis, whereas $q_x$ is reserved for the quasimomentum dependence of the single-particle SOC Hamiltonian.

The collective density excitations of the condensate act as two effective atomic oscillators with frequency $\Omega = \hbar k^2/m_{\mathrm{bec}}$, driven by radiation pressure. Linearizing Eqs.~(\ref{QLE}) (see \cite{supply}) yields the effective optomechanical coupling $G = \sqrt{2}\,g_a |c_s|$ and detuning $\Delta = \Delta_a + g_a N$, where 
$g_a = (\omega_c/L)\sqrt{\hbar / (m_{\mathrm{bec}} \Omega)}$ 
and $|c_s|$ is the steady-state intracavity amplitude. The quantity $m_{\mathrm{bec}} = \hbar \omega_c^2 / (L^2 U_0^2 \Omega)$ denotes the effective mass of the atomic mirror.

We extract topology directly from the measured cavity transmission spectrum. Using input–output theory on frequency domain solution of linearized quantum–Langevin equations, PSD of the output field, directly encodes the photonic Chern marker of the hybrid atom–cavity system~\cite{Silveirinha2019}, see supplementary materials \cite{supply} and Methods for details.
\begin{eqnarray}
	S_{\mathrm{out}}(P,\omega)
	= \Xi(\omega)\,E_{T,\omega}\,\frac{C_M(k_x,\omega)}{2\pi},
	\Xi(\omega)=\frac{2\,\eta_{\mathrm{det}}\,\kappa_{\mathrm{ext}}}{|R(\omega)|^{2}},
\end{eqnarray}
where $R(\omega)$ is the cavity response function and $E_{T,\omega}=\frac{\hbar\omega}{2}\coth\!\left(\frac{\hbar\omega}{2k_{B}T}\right)$. $C_M(k_x,\omega)$ is the local Chern marker carrying information bulk transport on edge modes, as illustrated in following findings.

\subsection*{Chiral-Photonic topology and Chern Marker}
\begin{figure}[htp]
	\centering
	\includegraphics[width=10cm]{Fig1.eps}
\caption{\textbf{Influence of SOC and interspecies interactions on the transmission PSD.}
	(\textbf{a})--(\textbf{c}) Transmitted-field PSD $S_{\rm out}(k_x,\omega)$ for increasing SOC strength
	$\alpha=1.5\,\Omega$ (\textbf{a}), $2.5\,\Omega$ (\textbf{b}), and $3.5\,\Omega$ (\textbf{c}), at fixed Raman and dissipation parameters (as in the main text).
	Larger $\alpha$ enhances spin--momentum locking, splits an almost degenerate response into two SOC-hybridized polaritonic branches, and strengthens the $\Omega_z$-controlled gap while increasing the $k$-asymmetry.
	(\textbf{d})--(\textbf{f}) PSD at fixed $\alpha$ for increasing interspecies interaction ratio $\epsilon\equiv U_{\uparrow\downarrow}/U$:
	$\epsilon=0$ (\textbf{d}), $\epsilon=1$ (\textbf{e}), and $\epsilon=2$ (\textbf{f}).
	Increasing $\epsilon$ renormalizes the atomic dispersion and dispersive light shift, shifting ridge frequencies and reducing gap contrast as the system approaches the SU(2)-symmetric point $\epsilon=1$.
	Across all panels, bright ridges trace poles of the linearized response, while linewidths reflect the net damping set by $\kappa$ and $\gamma$.}
	\label{Fig1}
\end{figure}
First, we consider the loss-dominated regime ($\kappa > \gamma$), where topology is trivial (Fig.~\ref{Fig1}). In Fig.~\ref{Fig1}(a), the pump interrogates the SOC-dressed pseudo-spin manifold. The transmitted field encodes the hybrid response in its power spectral density (PSD), $S_{\mathrm{out}}(P,\omega)$, where the ridges track the poles of the linear resolvent, and the linewidths are primarily determined by $\kappa$.

In Fig.~\ref{Fig1}(b), with the Raman (effective Zeeman) coupling switched off ($\Omega_z = 0$), the SOC does not imprint a $k$-resolved band structure on the optical readout. Consequently, the transmission reduces to a conventional cavity spectrum: two resolved, $\kappa$-broadened sidebands (Stokes/anti-Stokes-like quasiparticle features). The corresponding Chern marker in Fig.~\ref{Fig1}(d) vanishes across $(k_x,\omega)$, confirming the absence of edge spectral flow.

Turning on $\Omega_z = 3\Omega$, see Fig.~\ref{Fig1}(c), activates SOC hybridization and opens a Dirac-like gap. The upper and lower polaritonic bands separate around the former crossing, with a gap that scales with $\Omega_z$. Since photon loss exceeds atomic dissipation ($\kappa > \gamma$), no gap-spanning ridge emerges, and no mode connects the bands across $k$. The reconstructed Chern marker in Fig.~\ref{Fig1}(e) remains near zero across $(k_x,\omega)$, certifying a topologically trivial response.

Physically, $\Omega_z$ mixes the pseudo-spin states and opens a mass gap, while $\kappa > \gamma$ suppresses the non-Hermitian PT-symmetric  feedback required to amplify an edge channel or create spectral winding/exceptional points. Thus, the cavity acts as a bulk spectrometer: with $\Omega_z = 0$, it shows standard sidebands; with $\Omega_z \neq 0$, it resolves an SOC-induced gap. In both cases, the PSD-derived Chern marker quantitatively certifies trivial topology.
\begin{figure}[t]
	\centering
	\includegraphics[width=11cm]{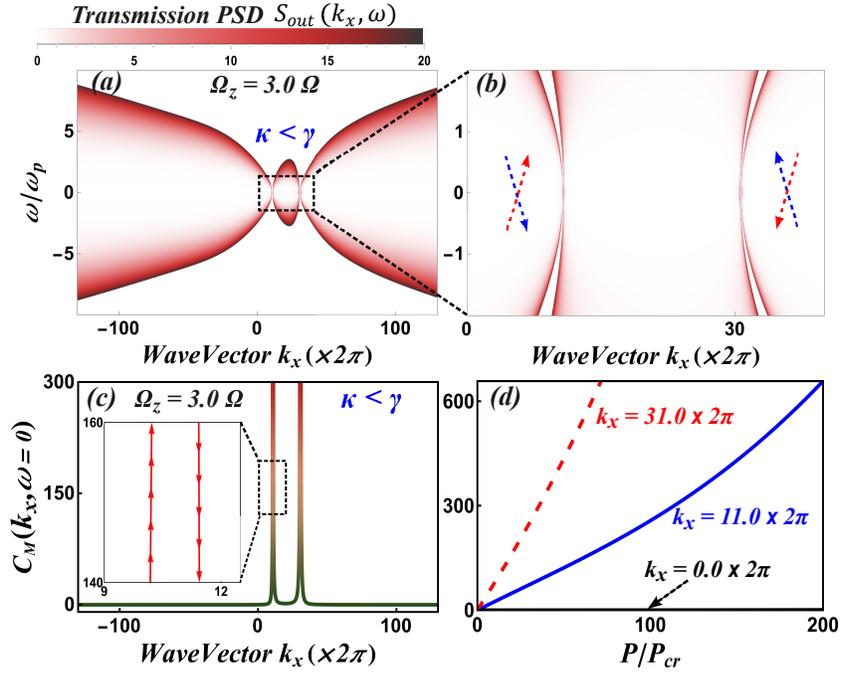}
	\caption{\textbf{Emergence of gap-spanning edge states and PSD-derived topology in the gain-dominated regime ($\gamma>\kappa$).}
		\textbf{a}, Power spectral density $S_{\mathrm{out}}(k_x,\omega)$ at Raman coupling $\Omega_{z}=3\,\Omega$ for $\gamma>\kappa$. A bright, continuous branch traverses the bulk gap, revealing an edge mode that connects the upper and lower polaritonic bands across wavevector $k$.
		\textbf{b}, Magnified view of the gap region in \textbf{a}. The inset arrows (blue/red) indicate opposite group velocities along the traversing branch, consistent with two-way chiral transport supported by non-Hermitian gain–loss imbalance.
		\textbf{c}, Chern-marker distribution reconstructed from the PSD in \textbf{a}. Two well-resolved positive peaks align with the gap-crossing branch and encode opposite transport directions of the chiral edge modes (signalled by the local spectral slope). The vanishing background away from the branch indicates that topology is concentrated at the gap-spanning trajectory.
	\textbf{d}, Chern marker as a function of input pump power $P$ for selected wavevectors $k_x$. The black curve ($k_x=0$) corresponds to the bulk region where no edge mode exists, giving a vanishing marker at all powers. The blue ($k_x=11\times2\pi$) and red ($k_x=31\times2\pi$) curves track the momenta of the first and second edge modes, respectively, both showing a pronounced increase of the marker with power as the chiral edge channels emerge and intensify with stronger light–matter coupling.}
	\label{Fig2}
\end{figure}

The principal result of this work is that the topology of a driven SOC–BEC in a single optical cavity can be read out \emph{spectroscopically} from the transmission PSD, without spatially resolving edges. In the gain-dominated regime ($\gamma > \kappa$), Fig.~\ref{Fig2}(a) shows that $S_{\mathrm{out}}(k, \omega)$ develops a bright, gap-spanning branch that connects the upper and lower polaritonic bands across the wavevector axis. This branch is the optical signature of an edge channel stabilized by the non-Hermitian imbalance: atomic dissipation feeding back through the cavity provides an effective amplification pathway that compensates photon leakage and enables a mode to thread the Raman-induced bulk gap. The enlarged view in Fig.~\ref{Fig2}(b) highlights the chiral nature of this transport: the inset arrows (blue/red) indicate opposite signs of the local spectral slope $\partial\omega/\partial k$, i.e., counter-propagating group velocities along the traversing ridge. This is expected when non-Hermitian coupling lifts reciprocity constraints and allows bidirectional chiral flow in frequency–wavevector space.

From the same transmission data, we reconstruct a frequency– and wavevector–resolved Chern marker, as shown in Fig.~\ref{Fig2}(c). Two narrow, positive lobes are pinned to the gap-spanning trajectory, reflecting the concentration of topological weight at the edge branch. Although the local marker is positive in both lobes (by construction of the PSD-based estimator), the opposite transport directions are encoded in the sign of the spectral slope and in the Berry-curvature distribution surrounding each lobe. Away from the edge trajectory, the marker collapses to zero, indicating that the bulk bands remain topologically inert while the edge mode carries the nontrivial winding. This one-to-one correspondence between a gap-crossing PSD ridge and a localized Chern-marker response establishes a direct, quantitative bridge between optical spectra and topology in a minimal (single-cavity) platform.

Figure~\ref{Fig2}(d) illustrates the dependence of the Chern marker on the cavity input power $P$ for specific wavevector values $k_x$, selected to correspond to distinct regions of the spectrum. The black curve, evaluated at $k_x = 0$, lies deep within the bulk band gap where no edge state exists; consequently, the Chern marker remains zero for all powers, confirming the absence of topological activity. The blue curve, taken at $k_x = 11 \times 2\pi$, corresponds to the momentum where the first edge mode emerges. Here, the Chern marker exhibits a sharp rise as the input power increases—signifying the onset of a topologically nontrivial regime once the optomechanical coupling $G \propto |c_s|$ becomes strong enough to overcome cavity losses and stabilize the chiral edge transport. The red curve, recorded at $k_x = 31 \times 2\pi$, traces the location of the second edge mode, which activates at slightly higher powers and yields a second distinct peak in the Chern marker. Together, these $k_x$-resolved curves reveal how the topological response can be tuned and selectively activated by controlling the pump power, mapping the successive appearance of edge channels directly onto an experimentally measurable photonic observable.

In the broader context of topological photonics, prior observations of edge transport typically relied on real-space imaging of waveguide arrays, photonic crystals, or ring-resonator lattices. In contrast, the present approach reads out edge physics from a \emph{bulk} transmission spectrum in a \emph{single} cavity, leveraging driven–dissipative (non-Hermitian) physics to generate a gap-spanning spectral branch and using a PSD-based Chern marker to certify its topology. This combination—gap-traversing edge ridges in $S_{\mathrm{out}}(k, \omega)$, chiral flow evidenced by opposite spectral slopes, and a co-localized Chern-marker signal with power-tunable strength—constitutes the core novelty of our protocol and establishes PSD spectroscopy as a compact, experimentally accessible route to topological diagnostics in hybrid light–matter systems.
\subsection*{Berry-curvature and non-Hermitian topology}
\begin{figure}[t]
	\centering
	\includegraphics[width=11cm]{Fig3.eps}
	\caption{\textbf{Berry-curvature cartography and non-Hermitian band topology.}
		\textbf{a}, Geometric spectral density $B(k_x,\omega)$ computed from the band eigenvectors and mode spectrum for $\Omega_{z}=3\,\Omega$, shown as a 3D surface with a base-plane density
		projection. Here $\omega$ denotes the analysis (Fourier) frequency of the transmitted field in the PSD measurement.
		The curvature weight concentrates along the gap-edge trajectories (spectral ridges) identified in transmission.
		\textbf{b}, Same as \textbf{a} for $\Omega_{z}=5\,\Omega$, revealing enhanced and more sharply localized curvature lobes
		and a redistribution of spectral weight as the Raman-induced gap increases.
		\textbf{c}, Real parts of the eigenvalues $\mathrm{Re}\,\lambda(k)$ at $\Omega_{z}=3\,\Omega$, displaying modal
		coalescence points characteristic of $\mathcal{PT}$-symmetric non-Hermitian band crossings.
		\textbf{d}, Imaginary parts $\mathrm{Im}\,\lambda(k)$ at the same parameters, showing linewidth bifurcation at the
		coalescence points in \textbf{c}, thereby identifying the exceptional points (EPs) that delimit the transition between
		unbroken and broken $\mathcal{PT}$ phases.}
	\label{Fig3}
\end{figure}

The geometric maps in Fig.~\ref{Fig3}(a,b) provide a momentum--frequency--resolved visualization of the band geometry
associated with the spectroscopic features observed in the cavity transmission. As described in the Methods section
\ref{mathods}, the quantity plotted in Fig.~\ref{Fig3}(a,b) is the geometric spectral density $B(k_x,\omega)$,
constructed by projecting the band-resolved geometric density onto the spectroscopic axis using the mode frequencies
$\omega_n(k_x)$ and linewidths $\Gamma_n(k_x)$. In this representation, $\omega$ is the analysis (Fourier) frequency of
the transmitted field and is therefore the natural axis for the experimentally measured spectral densities, while the
underlying geometry enters through the eigenvectors of the linearized dynamics.

For $\Omega_{z}=3\,\Omega$, Fig.~\ref{Fig3}(a) shows that $B(k_x,\omega)$ is narrowly concentrated along the gap-edge
trajectories, forming ridge-like lobes in the 3D surface and in the base-plane density projection. These ridges track
the same spectral loci where the transmission PSD exhibits maximal response, indicating that the measured gap-edge
features are accompanied by a pronounced concentration of band geometry. Increasing the Raman coupling to
$\Omega_{z}=5\,\Omega$ [Fig.~\ref{Fig3}(b)] sharpens these lobes and shifts the geometric weight outward in the
$(k_x,\omega)$ plane, consistent with an increased Raman-induced gap and a concomitant steepening of the dispersions.
Importantly, the geometric weight remains locked to the gap-edge trajectories: it is concentrated precisely where the
PSD develops a robust gap-spanning spectral branch. This behaviour provides a geometric counterpart to the peaks and
ridges highlighted by the PSD-derived Chern marker $C_M(k_x,\omega)$ defined in the Methods section \ref{mathods},
and it establishes a direct correspondence between the spectroscopic signature of spectral flow and the distribution of
band geometry in frequency space. Viewed across parameter sweeps, the evolution of $B(k_x,\omega)$ therefore offers an
intuitive ``cartography'' of how the underlying geometric response is organized in the spectrum of a driven
non-Hermitian light--matter system.

Figures~\ref{Fig3}(c,d) link this geometric structure to non-Hermitian band topology. At $\Omega_{z}=3\,\Omega$, the real
parts of the eigenvalues $\mathrm{Re}\,\lambda(k)$ in Fig.~\ref{Fig3}(c) exhibit modal coalescence points, while the
imaginary parts $\mathrm{Im}\,\lambda(k)$ in Fig.~\ref{Fig3}(d) display a simultaneous linewidth bifurcation at the same
parameters, identifying exceptional points (EPs) in the $\mathcal{PT}$-symmetric spectrum. These EPs demarcate the
transition between the unbroken phase (distinct mode frequencies with equal linewidths) and the broken phase (frequency
coalescence accompanied by asymmetric linewidths), and they occur proximate to the high-$B(k_x,\omega)$ ridges. Notably,
the EP locations inferred from $\mathrm{Re}\,\lambda(k)$ and $\mathrm{Im}\,\lambda(k)$ coincide with the $(k_x,\omega)$
region hosting the geometric hotspots and the gap-spanning spectral branch seen in transmission. This co-localization of
EPs, geometric concentration, and gap-traversing spectral flow indicates that the observed edge-like response in the
$\gamma>\kappa$ (strong atomic-dissipation) regime is governed by non-Hermitian criticality and its associated mode
hybridization. Together, the combined geometric mapping and EP spectroscopy in a single-mode cavity provide a compact
and experimentally accessible route to diagnosing band geometry and non-Hermitian topology in driven hybrid
light--matter systems, using only frequency-resolved transmission measurements augmented by eigenmode-based geometric
analysis.

\subsection*{Edge modes with asymmetric phase transition of SOC-BEC}
\begin{figure}[ht]
	\centering
	\includegraphics[width=15cm]{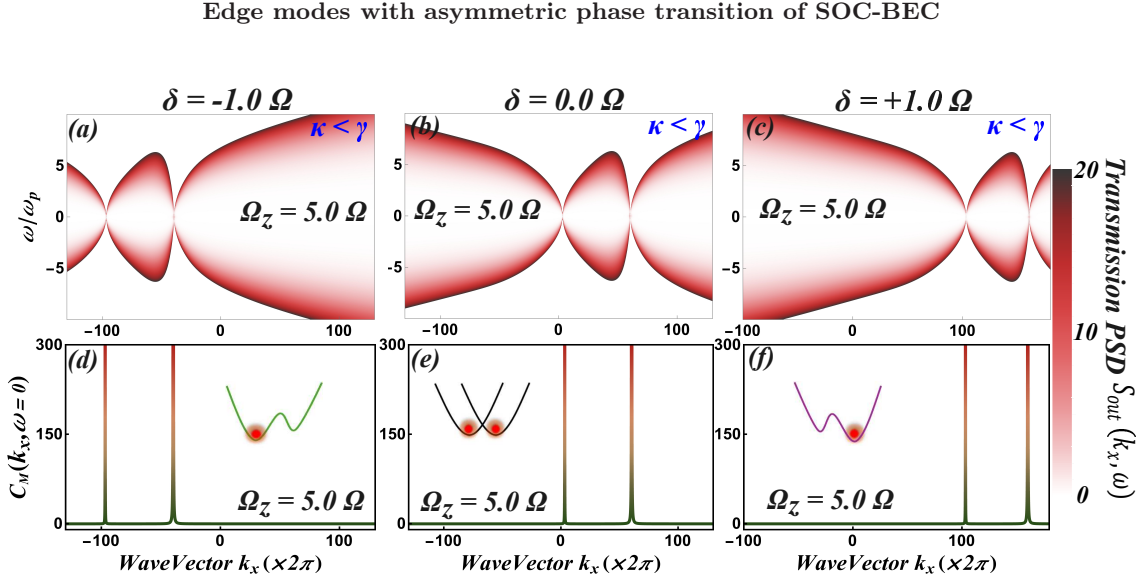}
	\caption{\textbf{Phase–detuning control of edge transport and Chern-marker localization.}
		Upper row: \textbf{a–c}, Transmission PSD $S_{\mathrm{out}}(k_x,\omega)$ for Raman detuning $\delta/\Omega=-1,\,0,\,+1$, respectively, at fixed $\Omega_z=5\,\Omega$ in the gain-dominated regime ($\gamma>\kappa$). A finite detuning breaks the $k_x\!\to\!-k_x$ symmetry of the SOC–BEC phase and \emph{re-positions} the gap-spanning edge branch within the bulk gap: for $\delta=-\Omega$ it is biased to negative $k_x$ (\textbf{a}), for $\delta=0$ it is centered (\textbf{b}), and for $\delta=+\Omega$ it shifts to positive $k_x$ (\textbf{c}). Lower row: \textbf{d–f}, Corresponding Chern-marker profiles along $k_x$ (frequency-integrated around the gap) showing a rigid displacement of the topological weight that tracks the edge-branch drift: left-shift for $\delta=-\Omega$ (\textbf{d}), symmetric/centered for $\delta=0$ (\textbf{e}), and right-shift for $\delta=+\Omega$ (\textbf{f}). The marker magnitude remains comparable, indicating that detuning redistributes rather than quenches topology, enabling \emph{phase-bias control} of chiral transport.}
	\label{Fig4}
\end{figure}
The data in Fig.~\ref{Fig4} establish Raman detuning $\delta$ as a precise \emph{phase-bias knob} that repositions chiral transport in momentum space without quenching topology. In the upper row, $S_{\mathrm{out}}(k_x,\omega)$ shows that a nonzero $\delta$ breaks the $k_x \to -k_x$ inversion of the SOC manifold and shifts the gap-spanning edge ridge: for $\delta = -\Omega$, as shown in Fig.~\ref{Fig4}(a), the traversing branch is biased to negative $k_x$; for $\delta = 0$, Fig.~\ref{Fig4}(b) shows the ridge centered; and for $\delta = +\Omega$, Fig.~\ref{Fig4}(c) shows the ridge shifted to positive $k_x$ \cite{Lin2011}. This drift is consistent with the SOC Hamiltonian used here, where $\delta$ acts as an effective Zeeman term that tilts the Dirac mass and displaces the spectral flow in $(k_x, \omega)$. Under gain-dominated conditions ($\gamma > \kappa$), the non-Hermitian imbalance then selects the shifted branch for amplification, yielding a bright gap-spanning ridge.

The lower row quantifies the same effect geometrically. Frequency-integrated Chern-marker profiles along $k_x$ [Figs.~\ref{Fig4}(d–f)] exhibit a rigid lateral translation that \emph{tracks} the PSD ridge: the peak of the marker migrates left for $\delta = -\Omega$, remains centered for $\delta = 0$, and moves right for $\delta = +\Omega$. Importantly, the marker amplitude remains nearly unchanged, indicating that detuning \emph{redistributes} topological weight in momentum space rather than diminishing it. To leading order, this behavior is expected from a curvature-conservation picture: the Berry curvature condenses near the gap-crossing trajectory, and $\delta$ shifts where the non-Hermitian spectral winding is maximal, but the integrated weight (set by the edge-channel topology) is preserved. 

Operationally, this realizes momentum-space routing of chiral transport in a single-mode cavity: by tuning $\delta$, one steers the $k_x$ at which the edge channel carries peak flow, providing a compact alternative to spatially engineered non-reciprocity in extended photonic lattices and $\mathcal{PT}$-symmetric platforms \cite{Baumann2010, harari2018, bandres2018, Bergholtz2021}.

From an experimental standpoint, the joint PSD/marker readout offers a direct calibration loop: the displacement of the PSD ridge with $\delta$ gives a spectroscopic handle on the effective Zeeman bias, while the co-moving Chern-marker peak certifies that the observed transport is genuinely topological rather than a dispersive artifact. The combination of phase-bias control (via $\delta$) and gain–loss selectivity (via $\gamma/\kappa$) thus enables deterministic placement and tuning of chiral edge channels in momentum space, leveraging non-Hermitian band topology in a minimal cavity-QED architecture.

\subsection*{Discussion}
We have demonstrated that cavity transmission spectroscopy offers a quantitative and experimentally accessible method for diagnosing band topology in a driven SOC–BEC, using a single cavity. By extracting a frequency- and wavevector-resolved Chern marker from the transmitted PSD, a standard optical observable is transformed into a topological probe. Gap-traversing ridges in $S_{\mathrm{out}}(k_x,\omega)$ co-localize with Chern-marker peaks when edge transport is present, while purely bulk spectra yield a vanishing marker, providing a direct optical signature of topological edge modes.

Three key findings underpin this conclusion. (i) \emph{Dissipation-driven topology:} In the loss-dominated regime ($\kappa > \gamma$), the Raman field opens a Dirac-like gap but no edge mode is stabilized, and the Chern marker remains zero. In the gain-dominated regime ($\gamma > \kappa$), a bright, gap-spanning ridge appears, with opposite spectral slopes (chiral flows) and a Chern marker concentrated along the trajectory of the edge mode. Power sweeps at selected $k_x$ values reveal the sequential activation of edge channels. (ii) \emph{Geometric and non-Hermitian certification:} The Berry curvature reconstructed from the PSD condenses along gap edges and intensifies with increased Raman coupling, while real and imaginary eigenvalue pairs reveal $\mathcal{PT}$-symmetric exceptional points that delineate the transition between unbroken and broken phases. (iii) \emph{Phase-bias control:} Raman detuning shifts the edge ridge and Chern marker across $k_x$ without reducing their amplitudes, enabling momentum-space routing of chiral transport within a minimal cavity-QED architecture.

These results establish cavity transmission PSD as a robust, quantitative probe of band topology in a driven SOC–BEC. The frequency- and wavevector-resolved Chern marker, extracted directly from $S_{\mathrm{out}}(P,\omega)$, co-localizes with gap-traversing ridges only in the gain-dominated regime ($\gamma > \kappa$), certifying the presence of edge transport. In the loss-dominated regime ($\kappa > \gamma$), the Chern marker vanishes, confirming the absence of edge states. Berry-curvature maps provide additional geometric insight, while $\mathcal{PT}$-symmetric eigenvalue coalescences identify exceptional points that govern the onset of chiral, gap-spanning transport.

The novelty of this work lies in using standard optical tools (PSD and Chern marker) to diagnose non-Hermitian topological features in a single-cavity system. This approach avoids the need for real-space imaging and instead relies on bulk transmission spectra to identify edge states and topological transport. The ability to tune and route chiral edge transport in momentum space via Raman detuning provides a compact and experimentally accessible method for controlling topological modes in driven light–matter systems.

In broader terms, this work fills a significant gap by demonstrating how non-Hermitian physics can be used to control and quantify topological transport in minimal cavity architectures, expanding the possibilities for future quantum photonic devices. Furthermore, this technique offers a pathway to studying geometric responses and non-Hermitian criticality in complex photonic systems with fewer experimental components compared to conventional lattice-based platforms. By extending this approach to multimode or Floquet cavities, we foresee the potential for developing spectroscopic topological photonics, where topological phases and edge channels can be engineered, read out, and tuned with standard optical methods, providing a powerful tool for quantum sensing, robust transport, and novel quantum optics applications.

\newpage

\section*{Methods}\label{mathods}
\subsection*{Extraction of the Chern Marker from Cavity Output Spectrum}
Topological features of the hybrid light--matter system manifest as robust gap-spanning spectral branches in the
fluctuations of the intracavity field. We access these fluctuations via the power spectral density (PSD) of the cavity
transmission, obtained from two-frequency correlation functions of the linearized quantum Langevin equations (QLEs).
For the driven cavity output field, input--output theory gives
\begin{equation}
	\hat{a}_{\mathrm{out}}(\omega) = \hat{a}_{\mathrm{in}}(\omega)
	-\sqrt{2\kappa_{\mathrm{ext}}}\,\hat{a}(\omega),
	\quad
	\hat{a}(\omega) = \tilde{\chi}_{c}(\omega)\,\hat{\mathcal{F}}(\omega),
\end{equation}
where $\kappa_{\mathrm{ext}}$ is the coupling rate to the detection port, $\hat{\mathcal{F}}$ collects optical and atomic
noise sources, and $\tilde{\chi}_{c}(\omega)=1/R(\omega)$ is the complex cavity response function obtained from the
linearized QLEs (explicit form given in the Supplementary derivation associated with Eq.~(\ref{PSD}), see \cite{supply}).
The resulting steady-state PSD of the output field reads
\begin{equation}\begin{split}
		S_\text{out}(P,\omega) &= \langle \hat{a}^{\dagger}_{\mathrm{out}}(\omega)\,
		\hat{a}_{\mathrm{out}}(\omega)\rangle\\
		&=\frac{2\pi}{|R(\omega)|^2}\Big([\kappa^2 + \omega^2 + \Delta^2 + 2\kappa\Delta]\\
		& \hspace{2.6cm} + 4\kappa\Delta\,[G S_{\uparrow}(\omega,\Delta) + G S_{\downarrow}(\omega,\Delta)]\Big).
	\end{split}\label{PSD}
\end{equation}

For later use we rewrite the PSD in a calibrated form,
\begin{equation}
	S_{\mathrm{out}}(P,\omega)
	= \frac{2\,\eta_{\mathrm{det}}\kappa_{\mathrm{ext}}}{|R(\omega)|^{2}}\,S_{\mathrm{edge}}(\omega),
	\qquad
	\Xi(\omega) = \frac{2\,\eta_{\mathrm{det}}\,\kappa_{\mathrm{ext}}}{|R(\omega)|^{2}},
	\label{eq:Sout_method}
\end{equation}
where $\eta_{\mathrm{det}}$ is the detection efficiency and $S_{\mathrm{edge}}(\omega)$ denotes the experimentally
inferred spectral contribution associated with the gap-spanning (edge-like) branch of the spectrum.
In a driven open system, $S_{\mathrm{edge}}(\omega)$ should be understood as an operational decomposition of the
	measured spectrum into a gap-spanning (edge-like) contribution versus bulk-like bands. It is not assumed to be an
	equilibrium thermal circulating-current spectrum.

In closed topological photonic cavities, thermal fluctuation spectra of circulating currents can be quantized in band gaps
and related to a global Chern number (Silveirinha~\cite{Silveirinha2019}). We use this result as motivation that
spectral densities can encode topology, while emphasizing that the present platform is driven and open. We therefore
define a momentum- and frequency-resolved \emph{Chern marker} $C_M(k_x,\omega)$ as an \emph{operational spectroscopic
	marker of gap-spanning spectral flow}, obtained by normalizing the calibrated PSD by an energy scale $E_{T,\omega}$:
\begin{equation}
	S_{\mathrm{out}}(P,\omega)
	= \Xi(\omega)\,E_{T,\omega}\,\frac{C_M(k_x,\omega)}{2\pi},
	\label{eq:final_mapping}
\end{equation}
or equivalently,
\begin{equation}
	C_M(k_x,\omega) \equiv \frac{2\pi}{\Xi(\omega)\,E_{T,\omega}}\,S_{\mathrm{out}}(P,\omega).
	\label{eq:Cdef}
\end{equation}
Here $E_{T,\omega}=\frac{\hbar\omega}{2}\coth\!\left(\frac{\hbar\omega}{2k_BT}\right)$ provides a convenient
quantum/thermal energy normalization of the spectral density. With this definition, $C_M(k_x,\omega)$ highlights the loci
in the $(k_x,\omega)$ maps where the spectrum develops a robust ridge that threads the Raman-induced gap. In the
topological regime this ridge persists under parameter variations and appears as a prominent hotspot in the PSD; in the
trivial regime the ridge is absent and $C_M(k_x,\omega)$ correspondingly lacks a gap-spanning feature.

Finally, the marker $C_M(k_x,\omega)$ remains well defined in the presence of dissipation and exceptional-point (EP)
physics because it is extracted directly from the measured response. In particular, when $\kappa>\gamma$
(cavity-loss-dominated), the PSD shows gapped polariton modes with a weak gap-spanning contribution, whereas when
$\gamma>\kappa$ (strong atomic-dissipation regime) non-Hermitian features (including EP-induced modal coalescence) can be
enhanced, accompanied by pronounced gap-spanning spectral branches in the PSD.

\subsection*{Relation Between the Chern Marker and Berry Curvature}

To connect the measured $(k_x,\omega)$ maps to band geometry, we compute geometric quantities from the eigenvectors of the
linearized dynamical matrix (equivalently, from the effective non-Hermitian mode matrix used to generate the dispersion
plots). Denoting right/left eigenvectors by $|u_{n}^{R}(k_x)\rangle$ and $\langle u_{n}^{L}(k_x)|$, the biorthogonal Berry
connection is
\begin{equation}
	\mathcal{A}_{n}(k_x)= i\langle u_{n}^{L}(k_x)|\nabla_{k_x}u_{n}^{R}(k_x)\rangle .
\end{equation}
For the one-dimensional momentum cut used in this work, the directly defined geometric invariant is the corresponding
Berry (Zak) phase along the Brillouin zone. Nevertheless, because the experiment probes frequency-resolved spectral
densities, it is convenient to visualize how the band geometry is distributed over the spectroscopic axis by introducing
an $\omega$-resolved geometric spectral density,
\begin{equation}
	B(k_x,\omega)=\sum_{n}\Omega_{n}(k_x)\,A_{n}(k_x,\omega),
	\label{eq:Bspec}
\end{equation}
where $\Omega_{n}(k_x)$ denotes the band-resolved geometric density computed from the eigenvectors (using the same
conventions as in the dispersion calculations), and $A_{n}(k_x,\omega)$ is a normalized spectral weight centered at the
mode frequency $\omega_{n}(k_x)$ with linewidth $\Gamma_{n}(k_x)$. A convenient choice is a Lorentzian,
\begin{equation}
	A_{n}(k_x,\omega)=\frac{1}{\pi}\frac{\Gamma_{n}(k_x)/2}{\left[\omega-\omega_{n}(k_x)\right]^{2}+\left[\Gamma_{n}(k_x)/2\right]^{2}},
	\label{eq:An}
\end{equation}
which reduces to a $\delta$-function on the band in the narrow-linewidth limit.

Since Berry curvature corresponds to a Chern-density (up to the factor $2\pi$) in conventional band topology, we define
the \emph{geometric Chern-marker distribution} by normalization,
\begin{equation}
	C_M^{(\mathrm{geom})}(k_x,\omega)\equiv \frac{B(k_x,\omega)}{2\pi},
	\qquad\text{so that}\qquad
	B(k_x,\omega)=2\pi\,C_M^{(\mathrm{geom})}(k_x,\omega).
	\label{eq:CMgeom}
\end{equation}
This definition provides a frequency-resolved representation of band geometry that can be compared directly with the
PSD-derived marker $C_M(k_x,\omega)$ obtained from Eq.~(\ref{eq:final_mapping}). Importantly, $\omega$ is the analysis
(Fourier) frequency of the transmitted field and is therefore a natural axis for spectral densities in photonics; it is
not treated as an independent crystal-momentum coordinate. In our analysis, agreement between $C_M(k_x,\omega)$ and
$C_M^{(\mathrm{geom})}(k_x,\omega)$ is assessed by their co-localization (and, where appropriate, their normalized
cross-correlation) on the same gap-spanning spectral ridge, rather than by assuming a pointwise identity.

\subsection*{Experimental Scheme for Chern Marker and Berry Curvature Measurement}

A spin--orbit-coupled Bose--Einstein condensate (SOC--BEC) is placed inside a high-finesse Fabry--P\'erot cavity and driven
to realize the hybrid atom--photon dynamics described in the main text. The power spectral density of the transmitted
cavity field is measured with a photodetector and frequency analyzer as system parameters (pump power, Raman coupling,
detuning, and the dissipation ratio $\gamma/\kappa$) are varied. Using the independently calibrated cavity response
$R(\omega)$ and detection factors, the momentum- and frequency-resolved spectroscopic Chern marker $C_M(k_x,\omega)$ is
obtained from the PSD via Eq.~(\ref{eq:final_mapping}) [or equivalently Eq.~(\ref{eq:Cdef})], yielding a direct
spectroscopic readout of gap-spanning spectral flow.

In parallel, the geometric Chern-marker distribution $C_M^{(\mathrm{geom})}(k_x,\omega)$ is computed from the eigenvectors
and mode frequencies $\omega_n(k_x)$ (including linewidths $\Gamma_n(k_x)$ where appropriate) using
Eqs.~(\ref{eq:Bspec})--(\ref{eq:CMgeom}). The co-localization of (i) the gap-spanning ridge in the measured PSD and (ii)
hotspots in $C_M^{(\mathrm{geom})}(k_x,\omega)$ provides a consistency check linking the spectroscopic marker
$C_M(k_x,\omega)$ to the underlying band geometry in the non-Hermitian hybrid system.

\subsection*{Effective Non-Hermitian Hamiltonian from the Quantum Langevin Formalism}
Although the microscopic dynamics of the cavity–BEC system are governed by a Hermitian Hamiltonian, the presence of photon leakage and atomic dissipation fundamentally alters its evolution, giving rise to an effective non-Hermitian description. This open-system nature is captured by the Lindblad master equation \cite{Lindblad1976,Breuer2002},

\begin{equation}
	\dot{\hat{\rho}} = -\frac{i}{\hbar}[\hat{\mathcal{H}},\hat{\rho}] + \sum_j \mathcal{D}[\hat{L}_j]\hat{\rho},
	\qquad 
	\mathcal{D}[\hat{L}]\hat{\rho} = \hat{L}\hat{\rho}\hat{L}^\dagger - \tfrac{1}{2}\{\hat{L}^\dagger \hat{L},\hat{\rho}\},
	\label{ME}
\end{equation}

where $\hat{\mathcal{H}}$ is the system Hamiltonian and $\mathcal{D}[\hat{L}_j]\hat{\rho}$ accounts for irreversible coupling to external reservoirs. For the cavity–atom platform, the relevant quantum jump operators are

\begin{equation}
\hat{L}_c = \sqrt{\kappa}\,\hat{c}, 
\hat{L}_{a,\sigma}(\mathbf r) = \sqrt{\gamma}\,\hat{\psi}_\sigma(\mathbf{r}),
\end{equation}

which model photon loss from the cavity field and spontaneous emission from the atomic modes, with decay rates $\kappa$ and $\gamma$, respectively.

After inserting these operators into Eq.~\eqref{ME} and separating reversible and dissipative contributions, the dynamics may be rewritten as

\begin{equation}
	\dot{\hat{\rho}} = -\frac{i}{\hbar}\bigl(\hat{H}_{\mathrm{eff}}\hat{\rho} - \hat{\rho}\hat{H}_{\mathrm{eff}}^\dagger\bigr) + \sum_j \hat{L}_j \hat{\rho} \hat{L}_j^\dagger,
	\label{ME_Heff}
\end{equation}

where the non-Hermitian operator governing the coherent (no-jump) evolution is

\begin{align}
	\hat{H}_{\mathrm{eff}} 
	&= \hat{\mathcal{H}}
	- \frac{i\hbar}{2}\sum_j \hat{L}_j^\dagger \hat{L}_j \nonumber \\
	&= \hat{\mathcal{H}} 
	- \frac{i\hbar}{2} 
	\left[
	\kappa\,\hat c^\dagger \hat c
	+ \gamma \!\int d\mathbf{r}\sum_\sigma 
	\hat{\psi}_\sigma^\dagger(\mathbf r)\hat{\psi}_\sigma(\mathbf r)
	\right].
	\label{Heff}
\end{align}

This complex Hamiltonian generates evolution with a decaying norm and corresponds to the "no-quantum-jump" part of the quantum trajectory picture \cite{Dalibard1992,Plenio1998,Carmichael1993}, while the stochastic terms $\hat{L}_j \hat{\rho} \hat{L}_j^\dagger$ account for sudden emission events that restore probability. The imaginary terms in Eq.~\eqref{Heff} therefore encode the irreversible loss of excitations and are responsible for non-Hermitian spectral properties such as complex eigenfrequencies, mode coalescence, and linewidth asymmetry.

The same structure emerges in the Heisenberg picture. Eliminating environmental degrees of freedom in the Born–Markov approximation \cite{Gardiner1985,WallsMilburn1994} yields the quantum Langevin equations

\begin{align}
	\dot{\hat c}(t) &= -\frac{i}{\hbar}[\hat c, \hat{\mathcal{H}}] - \frac{\kappa}{2}\hat c(t) + \sqrt{\kappa}\,\hat c_{\mathrm{in}}(t), \label{QLE_c}\\
	\dot{\hat\psi}_\sigma(\mathbf r,t) &= -\frac{i}{\hbar}[\hat\psi_\sigma(\mathbf r),\hat{\mathcal{H}}] 
	- \frac{\gamma}{2}\hat\psi_\sigma(\mathbf r,t) 
	+ \sqrt{\gamma}\,\hat\psi_{\sigma,\mathrm{in}}(\mathbf r,t). \label{QLE_a}
\end{align}

Neglecting the input noise operators $\hat c_{\mathrm{in}}$ and $\hat\psi_{\mathrm{in}}$, Eqs.~\eqref{QLE_c}–\eqref{QLE_a} are equivalent to Heisenberg evolution under $\hat{H}_{\mathrm{eff}}$. Thus, the Langevin and Lindblad approaches are fully consistent: the former separates coherent decay from fluctuating quantum noise, while the latter embeds both in a single master equation.

To analyze small fluctuations and spectral topology, we linearize the dynamics by expanding each operator around its steady state, $\hat c = c_s + \delta\hat c$ and $\hat\psi = \psi_s + \delta\hat\psi$. Collecting all fluctuations in the vector $\delta\hat{\boldsymbol{X}}$, the equations of motion take the form

\begin{equation}
	\frac{d}{dt}\,\delta\hat{\boldsymbol{X}}(t) = K\,\delta\hat{\boldsymbol{X}}(t) + \hat{\boldsymbol{\xi}}(t),
	\label{drift}
\end{equation}

where $\hat{\boldsymbol{\xi}}$ contains the input noise terms and $K$ is the linearized drift matrix derived from $\hat{H}_{\mathrm{eff}}$. This matrix can be written as

\begin{equation}
K = K_{\mathrm{H}} + K_{\mathrm{D}}, \qquad
K_{\mathrm{H}}^\dagger = -K_{\mathrm{H}}, \quad
K_{\mathrm{D}} = -\mathrm{diag}\!\left(\tfrac{\kappa}{2},\tfrac{\gamma}{2},\dots\right),
\end{equation}

making explicit its decomposition into coherent (anti-Hermitian) and dissipative (Hermitian) parts.

The eigenvalues of $K$,

\begin{equation}
\lambda_n = \omega_n - i\,\Gamma_n/2,
\end{equation}

encode both the oscillation frequencies $\omega_n$ and damping rates $\Gamma_n$ of the coupled excitations. Their distribution in the complex plane determines dynamical stability (all $\mathrm{Re}[\lambda_n]<0$), exceptional points, linewidth anisotropies, and topological winding of the spectrum. The associated resolvent

\begin{equation}
G^R(\omega) = \big[-i\omega\mathbb{1} - K\big]^{-1}
\end{equation}

provides access to observable quantities such as the cavity transmission, power spectral density, and dynamical susceptibility. In particular, exceptional-point physics and non-Hermitian topological features arise when decay imbalance ($\gamma > \kappa$) drives eigenvalue coalescence or induces complex-energy winding in parameter space.

Hence, although the underlying microscopic Hamiltonian is Hermitian, the combination of cavity photon leakage and atomic dissipation leads naturally to a non-Hermitian framework. Whether approached via the Lindblad master equation, quantum trajectories, or the Langevin formulation, the result is the same: the effective Hamiltonian $\hat{H}_{\mathrm{eff}}$ and its linearized drift matrix $K$ provide a unified basis for describing spectral topology, non-Hermitian band structures, and dynamical instabilities in the SOC–BEC cavity system.

\section*{Data availability}
All data obtained and illustrated as finding of this work are computed from analytical expressions and solutions available within the article and its Supplemental
Information.

\begin{acknowledgments}
	K.A.Y. acknowledges the support of Research Fund for International Young Scientists by NSFC under grant No. KYZ04Y22050, Zhejiang Normal University research funding under grant No. ZC304021914 and Zhejiang province postdoctoral research project under grant number ZC304021952. W.M.L. acknowledges the support from National Key R\&D Program of China under grants No. 2021YFA1400900, 2021YFA0718300, 2021YFA1402100, NSFC under Grants Nos. 12174461, 12234012, 12334012, 52327808, Space Application System of China Manned Space Program.
\end{acknowledgments}

\section*{Author contributions}
The research was initiated and led by K.A.Y.;  K.A.Y. and G.X.L. analyzed the results; K.A.Y wrote the paper under guidance G.X.L and W.M.L.

\section*{Code availability}
All results illustrated in this word are generated from the known analytical expressions given in the article and its Supplemental
Information. The detailed code is available from the corresponding author upon reasonable request.

\section*{Competing interests}
The authors declare no competing interests.

\end{document}


\title{Supplemental Material : Spectroscopic readout of chiral photonic topology in a single-cavity spin-orbit-coupled Bose–Einstein condensate}
\author{Kashif Ammar Yasir}
\email {kayasir@zjnu.edu.cn}\affiliation{Department of Physics, Zhejiang Normal University, Jinhua 321004, China.}
\affiliation{Zhejiang Institute of Photoelectronics, Jinhua 321004, China.}
\author{Gao Xianlong}
\email {gaoxl@zjnu.edu.cn}\affiliation{Department of Physics, Zhejiang Normal University, Jinhua 321004, China.}

\date{\today}
\maketitle
\subsection*{Linearized Quantum Langevin Equations}

The full quantum dynamics of the cavity–BEC system are governed by nonlinear quantum Langevin equations (QLEs), which include operator products and noise terms arising from dissipation. Under strong coherent driving, the system reaches a steady state around which quantum fluctuations remain small. In this regime, we linearize the QLEs by splitting each operator into its steady-state mean value and a small fluctuation,
$
\hat{\mathcal{O}}(t) = \mathcal{O}_{s} + \delta\hat{\mathcal{O}}(t),
$
where $\mathcal{O}_{s}$ is the classical steady-state component, and $\delta\hat{\mathcal{O}}(t)$ captures the residual quantum fluctuations. The linearization is valid when the intracavity photon number is large, such that higher-order fluctuation terms may be safely neglected~\cite{Yasir_2022,Yasir2017}.

For concreteness, we assume that the two pseudo-spin components of the condensate share equal populations at equilibrium, 
$\hat{\varphi}_\uparrow^{\dagger}\hat{\varphi}_\uparrow
= \hat{\varphi}_\downarrow^{\dagger}\hat{\varphi}_\downarrow = N/2$.
To express the linearized dynamics in a compact form, we introduce dimensionless field quadratures
$
\hat{q}_O = \frac{1}{\sqrt{2}}(\hat{O} + \hat{O}^\dagger),
\hat{p}_O = \frac{i}{\sqrt{2}}(\hat{O}^\dagger - \hat{O}),
$
which fulfil $[\hat{q}_O,\hat{p}_O] = i$, corresponding to $\hbar = 1$. These quadratures represent amplitude- and phase-type fluctuations and form a convenient basis for linear stability and spectral analyses.

The linearized equations of motion take the matrix form
$
\dot{\mathcal{X}} = \mathcal{K}\mathcal{X} + \mathcal{F},
$
where $\mathcal{X} = [\delta q_c, \delta p_c, \delta q_\uparrow, \delta p_\uparrow, \delta q_\downarrow, \delta p_\downarrow]^{T}$ collects all fluctuation quadratures. The corresponding noise vector is
$\mathcal{F} = [\sqrt{2\kappa}\, q_c^{\mathrm{in}}, \sqrt{2\kappa}\, p_c^{\mathrm{in}}, 0, 2\sqrt{2\gamma} f_a, 0, 2\sqrt{2\gamma} f_a]^{T}$.
The drift matrix $\mathcal{K}$ that governs the dynamics is
\begin{widetext}
	\begin{center}
		\[
		\mathcal{K}=
		\begin{pmatrix}
			-\kappa & \Delta & 0 & 0 & 0 & 0 \\
			\Delta & -\kappa & G_a & 0 & G_a & 0 \\
			2G_a & 0 & M & \tfrac{\Omega_z}{2} & \alpha - \tfrac{\delta}{2} & 0 \\
			0 & 0 & \tfrac{\Omega_z}{2} & M & 0 & -(\alpha - \tfrac{\delta}{2}) \\
			2G_a & 0 & -\alpha + \tfrac{\delta}{2} & 0 & M & -\tfrac{\Omega_z}{2} \\
			0 & 0 & 0 & \alpha - \tfrac{\delta}{2} & -\tfrac{\Omega_z}{2} & M
		\end{pmatrix}.
		\]
	\end{center}
\end{widetext}

Here,
$M = \frac{\Omega}{2} + v + U N (1 - \varepsilon) - \gamma$
is the effective atomic damping, which incorporates recoil $\Omega = \hbar k^{2}/m_a$, interaction shifts, and SOC corrections. The enhanced couplings
$G_{a} = \sqrt{2}\, g_{a} |c_s|$ depend on the steady-state intracavity field $c_{s}$. The effective detuning is $\Delta = \tilde{\Delta}+ g_{a}N$, including both mechanical displacement and dispersive coupling to atoms.

\subsection*{Routh--Hurwitz Stability Criterion} \label{rt}
To ensure that the steady-state solution of the linearized system is dynamically stable, we analyze the drift matrix $\mathcal{K}$ using the Routh--Hurwitz criterion. A steady state is stable only if all eigenvalues of $\mathcal{K}$ have strictly negative real parts. Instead of computing these eigenvalues explicitly, one can determine stability by examining the characteristic polynomial
\[
p(s) = \det(s\mathbb{1} - \mathcal{K}) 
= s^{6} + a_{1}s^{5} + a_{2}s^{4} + a_{3}s^{3} 
+ a_{4}s^{2} + a_{5}s + a_{6}.
\]
The coefficients $a_{j}$ depend on the system parameters $\kappa$, $\gamma$, $M$, $\Delta$, $\Omega_{z}$, $\alpha$, $\delta$, and $G_{a}$. A symbolic expansion of $p(s)$ yields compact forms for the first few coefficients, which already encode essential stability requirements:

\begin{align}
	a_{1} &= 2(\kappa - 2M), \\[2pt]
	a_{2} &= \kappa^{2} - \Delta^{2} + 6M^{2} - 8M\kappa 
	- \tfrac{1}{2}\Omega_{z}^{2} + 2\alpha^{2} 
	- 2\alpha\delta + \tfrac{1}{2}\delta^{2}, \\[2pt]
	a_{3} &= 4\Delta^{2}M - 4\Delta G_{a}^{2} - 4M^{3} + 12M^{2}\kappa
	+ M\Omega_{z}^{2}- 4M\alpha^{2} \nonumber\\
	& + 4M\alpha\delta - M\delta^{2}\quad - \Omega_{z}^{2}\kappa + 4\alpha^{2}\kappa
	- 4\alpha\delta\,\kappa + \delta^{2}\kappa .
\end{align}
The Routh--Hurwitz criterion states that all roots of $p(s)$ lie in the left half-plane (i.e., $\mathrm{Re}[s] < 0$) if and only if all principal Hurwitz determinants $\Delta_{j}$ are positive \cite{Yasir_2022,Yasir2017}:
\[
\Delta_{1} > 0,\ \Delta_{2} > 0,\ \Delta_{3} > 0,\ \Delta_{4} > 0,\ 
\Delta_{5} > 0,\ \Delta_{6} > 0.
\]
For a sixth-order polynomial, the first three determinants can be written explicitly in terms of the coefficients:
\begin{align}
	\Delta_{1} &= a_{1} > 0, \label{RH1} \\[2pt]
	\Delta_{2} &= a_{1}a_{2} - a_{3} > 0, \label{RH2} \\[2pt]
	\Delta_{3} &= a_{1}a_{2}a_{3} - a_{1}^{2}a_{4} - a_{3}^{2} + a_{1}a_{5} > 0. \label{RH3}
\end{align}
The higher-order determinants $(\Delta_{4}, \Delta_{5}, \Delta_{6})$ are more cumbersome but can be evaluated numerically from the polynomial coefficients for any chosen set of system parameters.

\medskip
\noindent\textbf{Compact stability requirements.}  
For practical implementation, the necessary and sufficient stability conditions can be summarized as:
\begin{widetext}
	\begin{equation}\label{eq:RHcompact}
		\begin{aligned}
			& a_{1} = 2(\kappa - 2M) > 0 
			&& \Rightarrow\ \kappa > 2M, \\[3pt]
			& a_{2} > 0 
			&& \Rightarrow\ 
			\kappa^{2} - \Delta^{2} - \tfrac{1}{2}\Omega_{z}^{2}
			+ 2(\alpha^{2} - \alpha\delta) + \tfrac{1}{2}\delta^{2}
			+ 6M^{2} - 8\kappa M > 0, \\[3pt]
			& a_{3} > 0,\quad 
			\Delta_{2} = a_{1}a_{2} - a_{3} > 0,\quad
			\Delta_{3} > 0, \\[3pt]
			& a_{6} = \det(\mathcal{K}) > 0,\quad 
			\Delta_{4} > 0,\ \Delta_{5} > 0,\ \Delta_{6} > 0.
		\end{aligned}
	\end{equation}
\end{widetext}
The first inequality, $\kappa > 2M$, provides a simple physical interpretation: the effective atomic damping $M$ must remain sufficiently small compared with the cavity loss rate. The second condition shows how detuning, SOC, and Zeeman splitting contribute to stabilizing the system against the coherent atom–cavity coupling $G_{a}$. All numerical simulations presented in this work satisfy these Routh--Hurwitz constraints to ensure operation in a stable, steady-state regime \cite{yasir2023,yasir2024,yasir2025,Ref49}.

\subsection*{Fourier Domain Solutions and Power Spectral Density}
To analyze the fluctuation spectra, we transform the linearized Langevin equations into the frequency domain. This allows us to obtain analytical expressions for the cavity-field quadratures \cite{Yasir2017}. The intracavity position and momentum quadratures are given by
\begin{align}
	\delta q_{c}(\omega) &= \frac{1}{R(\omega)} \Big[ \sqrt{2\kappa} \big( \Delta\, \delta p_{c}^{\mathrm{in}} 
	+ (\kappa + i\omega)\delta q_{c}^{\mathrm{in}} \big) 
	+ \Delta G \big( \delta q_{\uparrow}(\omega) + \delta q_{\downarrow}(\omega) \big) \Big], \notag \\
	\delta p_{c}(\omega) &= \frac{1}{R(\omega)} \Big[ \sqrt{2\kappa} \big( \Delta\, \delta p_{c}^{\mathrm{in}} 
	+ (\kappa + i\omega)\delta q_{c}^{\mathrm{in}} \big) 
	+ (\kappa + i\omega)G \big( \delta q_{\uparrow}(\omega) + \delta q_{\downarrow}(\omega) \big) \Big].
\end{align}

Similarly, the position quadrature of the atomic spin components is
\begin{equation}
	\begin{split}
		\delta q_{\uparrow,\downarrow}(\omega) = \frac{1}{X(\omega)} \Big[ 
		(Z_{\uparrow,\downarrow}(\omega) + Y_{\uparrow,\downarrow}(\omega))\,C(\omega)\,
		\big(\Delta \delta p_{c}^{\mathrm{in}} + (\kappa + i\omega)\delta q_{c}^{\mathrm{in}} \big)
		+ F(\omega)\, f_{a} \Big].
	\end{split}
\end{equation}

In these expressions,
\[
R(\omega) = (\kappa + i\omega)^2 - \Delta^{2}, \qquad
D(\omega) = -\omega^{2} \big[ (\kappa + i\omega)^2 - \Delta^{2} \big],
\]
and atomic interactions introduce
\[
\Gamma(\omega) = \gamma_{a} + i\omega - \frac{\Omega}{2} - \nu - U N (1 - \varepsilon), \qquad
K(\omega) = M^{2}(\omega) + \big(\alpha^{2} - \tfrac{\delta}{2} \big)^{2}.
\]

The modified susceptibilities of the atomic modes are then
\begin{align}
	Y_{\uparrow,\downarrow}(\omega) &= 4M(\omega)K(\omega)R(\omega) 
	\pm \Omega_{z}^{2}R(\omega)D(\omega)
	- 8G^{2}\Delta K(\omega)D(\omega), \notag \\
	Z_{\uparrow,\downarrow}(\omega) &= \pm\Omega_{z}^{2}R(\omega)D(\omega)
	+ 4\big( \pm\alpha \mp \tfrac{\delta}{2} \big) K(\omega)R(\omega)D(\omega)
	+ 8G^{2}\Delta K(\omega)D(\omega), \notag \\
	C(\omega) &= 8G^{2} \sqrt{2\kappa} \, K(\omega)D(\omega), \notag \\
	F(\omega) &= \big[ Z_{\uparrow,\downarrow}(\omega) + Y_{\downarrow,\uparrow}(\omega) \big]\,
	8G^{2}\sqrt{\gamma_{a}}\,K(\omega)D(\omega), \notag \\
	X(\omega) &= Y_{\uparrow}(\omega)Y_{\downarrow}(\omega) 
	+ Z_{\uparrow}(\omega)Z_{\downarrow}(\omega).
\end{align}

\noindent
Using the standard two-frequency autocorrelation formalism~\cite{Yasir_2022,Ref49}, the power spectral density (PSD) of the atomic spin-up and spin-down modes is obtained as
\begin{equation}
	\begin{split}
		S_{\uparrow,\downarrow}(\omega,\Delta) 
		= \frac{1}{|X(\omega)|^{2}} \Bigg[
		2\pi |C(\omega)|^{2} \big( |Z_{\uparrow,\downarrow}(\omega)|^{2} 
		+ |Y_{\downarrow,\uparrow}(\omega)|^{2} \big) (\Delta^{2} + \kappa^{2} + \omega^{2})
		+ 2\pi F(\omega) \Bigg].
	\end{split}
\end{equation}

For later use, we express the linearized light--matter coupling as
$G=\sqrt{2}\,g_a |c_s|$, where the steady-state intracavity field is
$c_s=\eta/(\kappa+i\Delta)$ and the drive amplitude is
$|\eta|=\sqrt{P\kappa/(\hbar\omega_p)}$.
Hence,
\[
G=\sqrt{2}\,g_a\,\frac{\sqrt{P\kappa/(\hbar\omega_p)}}{\sqrt{\kappa^2+\Delta^2}} .
\]

To connect intracavity fields with the detected output, we apply the input–output relations
\[
\delta q_{c}^{\mathrm{out}} = \sqrt{2\kappa}\,\delta q_{c} - \delta q_{c}^{\mathrm{in}}, \qquad
\delta p_{c}^{\mathrm{out}} = \sqrt{2\kappa}\,\delta p_{c} - \delta p_{c}^{\mathrm{in}}.
\]
From this, the output quadratures are
\begin{align}
	\delta q_{c}^{\mathrm{out}}(\omega) &= \frac{1}{R(\omega)} \Big[ 
	2\kappa\Delta\,\delta p_{c}^{\mathrm{in}} 
	+ (\kappa^{2} + \omega^{2} + \Delta^{2})\delta q_{c}^{\mathrm{in}}
	+ \sqrt{2\kappa} \Delta G \big( \delta q_{\uparrow}(\omega) + \delta q_{\downarrow}(\omega) \big) \Big], \notag \\
	\delta p_{c}^{\mathrm{out}}(\omega) &= \frac{1}{R(\omega)} \Big[
	2\kappa\Delta\,\delta q_{c}^{\mathrm{in}} 
	+ (\kappa^{2} + \omega^{2} + \Delta^{2})\delta p_{c}^{\mathrm{in}} 
	+ \sqrt{2\kappa}(\kappa + i\omega)G\big( \delta q_{\uparrow}(\omega) + \delta q_{\downarrow}(\omega) \big) \Big]. 
\end{align}

Finally, the output field operator is obtained by recombining quadratures:
\begin{equation}
	\begin{split}
		\delta c_{\mathrm{out}}(\omega) 
		= \frac{1}{R(\omega)} \Big[
		2\kappa\Delta\,\delta c_{\mathrm{in}}^{\dagger}
		+ (\kappa^{2} + \omega^{2} + \Delta^{2})\delta c_{\mathrm{in}}
		+ \sqrt{2\kappa} \Delta G \big( \delta q_{\uparrow}(\omega) + \delta q_{\downarrow}(\omega) \big) \Big].
	\end{split}
\end{equation}

\begin{figure*}[tp]
	\centering
	\includegraphics[width=15cm]{SFig1.eps}
	\caption{\textbf{Influence of spin–orbit coupling and interspecies interactions on the transmission PSD.}
		\textbf{a–c}, Transmitted-field PSD \(S_{\mathrm{out}}(k,\omega)\) for increasing SOC strength:
		\(\alpha=1.5\,\Omega\) (a), \(\alpha=2.5\,\Omega\) (b), \(\alpha=3.5\,\Omega\) (c), at fixed Raman and dissipation parameters (as in the main text).
		Larger \(\alpha\) enhances spin–momentum locking, splitting a near-degenerate response into two SOC-hybridized polaritonic branches and opening/strengthening an \(\Omega_{z}\)-controlled gap with increasing \(k\)-asymmetry.
		\textbf{d–f}, PSD at fixed \(\alpha\) for increasing interspecies interaction ratio \(\varepsilon\equiv U_{\uparrow\downarrow}/U\):
		\(\varepsilon=0\,U\) (d), \(\varepsilon=1\,U\) (e), \(\varepsilon=2\,U\) (f).
		Growing \(\varepsilon\) renormalizes the atomic dispersion and dispersive light shift, shifting ridge frequencies and reducing gap contrast as the spin channels approach the SU(2)-symmetric point (\(\varepsilon=1\)). Across all panels, bright ridges trace poles of the linearized response, while linewidths reflect the net damping set by \(\kappa\) and \(\gamma\).}
	\label{SFig2}
\end{figure*}
\begin{figure*}[tp]
	\centering
	\includegraphics[width=10cm]{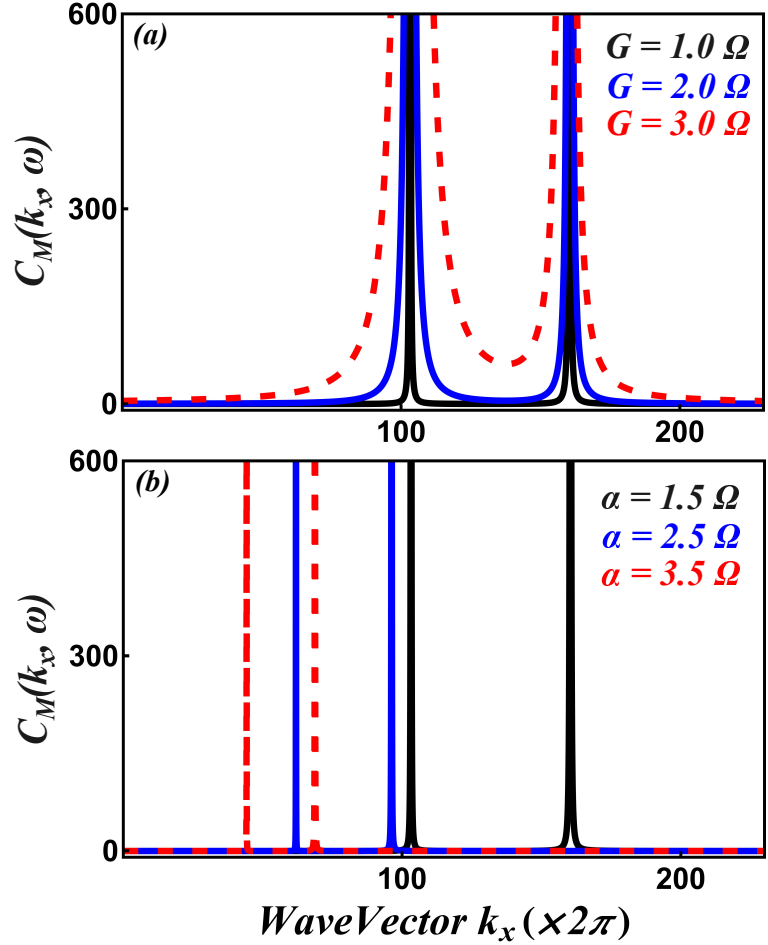}
	\caption{\textbf{Atom–cavity coupling and SOC control of the PSD-derived Chern marker.}
		\textbf{a}, Dependence of the photonic Chern marker (extracted from the transmission PSD) on the light–matter coupling strength \(G\): 
		\(G=1\,\Omega\) (black), \(G=2\,\Omega\) (blue), \(G=3\,\Omega\) (red).
		Increasing \(G\) enhances hybridization and cooperativity, amplifying the edge contribution to the marker and reducing the activation threshold set by losses.
		\textbf{b}, Dependence on the spin–orbit–coupling strength \(\alpha\): 
		\(\alpha=1.5\,\Omega\) (black), \(\alpha=2.5\,\Omega\) (blue), \(\alpha=3.5\,\Omega\) (red).
		Larger \(\alpha\) sharpens the SOC-induced band separation and redistributes the topological weight in \((k,\omega)\), shifting and strengthening the Chern-marker peaks co-localized with the gap-traversing edge ridge.}
	\label{SFig3}
\end{figure*}
\section{Influence of SOC and Interspecies Interactions on Transmission PSD}
Supplementary Fig.~\ref{SFig2} clarifies how single-particle spin–orbit coupling (SOC) and two-body collisions reorganize the photonic spectrum recorded in the transmission PSD. In Figs.~\ref{SFig2}(a–c), increasing \(\alpha\) from \(1.5\,\Omega\) to \(3.5\,\Omega\) strengthens the \(\alpha k_x \sigma_y\) term in the single-particle Hamiltonian, enhancing spin–momentum locking and the Raman-induced mixing of the pseudo-spin branches. Spectroscopically, the bright PSD ridges (which track the poles of the linearized resolvent) evolve from a near-degenerate sideband-like response to two well-separated hybrid polaritonic branches, with a Dirac-like gap set by \(\Omega_z\) and a growing left–right \(k\)-asymmetry that reflects the odd-in-\(k_x\) SOC coupling. At fixed \((\kappa, \gamma)\), the linewidths change only mildly with \(\alpha\), but the relative ridge intensities redistribute because the intracavity field couples more efficiently to spin-balanced superpositions as \(\alpha\) increases. Thus, \(\alpha\) serves as a “band-geometry” knob: it determines where spectral weight resides in \((k, \omega)\) space and how strongly the atomic manifold imprints its dispersion on the optical readout.

In Figs.~\ref{SFig2}(d–f), \(\alpha\) is held fixed, and the interspecies interaction ratio \(\varepsilon \equiv U_{\uparrow \downarrow} / U\) is varied from \(0\,U\) to \(2\,U\). Collisions modify the atomic self-energies and the mean dispersive light shift \(v \propto g_a n_s\), shifting the hybrid-mode frequencies and modulating their contrast. As \(\varepsilon\) approaches the SU(2)-symmetric point (\(\varepsilon = 1\), i.e., \(U_{\uparrow \downarrow} = U\)), the two spin channels become less distinguishable: the SOC-hybridized branches draw closer, and the apparent gap contrast is reduced, consistent with a diminished spin imbalance in the polaritonic eigenvectors. For \(\varepsilon > 1\), enhanced interspecies repulsion further renormalizes the dressed dispersions, producing additional ridge shifts and modest broadening due to stronger matter-mediated backaction (at fixed \(\kappa, \gamma\)). 

Overall, these trends clearly separate the roles of \(\alpha\) and \(\varepsilon\): \(\alpha\) primarily controls the **splitting and \(k\)-asymmetry** of the PSD bands (band geometry), while \(\varepsilon\) tunes the **interaction-induced renormalization and visibility** of the gap (spectral placement and contrast) without fundamentally altering the linewidths set by loss.

\section{Influence of Atom-Cavity Coupling and SOC on Chern Marker}
Supplementary Fig.~\ref{SFig3}(a) shows that the PSD-derived Chern marker is an increasing function of the atom–cavity coupling \(G\). As \(G\) is raised from \(1\,\Omega\) to \(3\,\Omega\), the effective cooperativity \(C \propto G^{2}/(\kappa\gamma)\) grows, reinforcing the light–matter hybridization that sustains the gap-spanning edge channel in the gain-dominated regime. Spectroscopically, the traversing ridge in \(S_{\mathrm{out}}(k,\omega)\) becomes brighter and more continuous, and the corresponding marker peak increases in magnitude, with its onset shifting towards weaker drive (or equivalently, tolerating larger damping). This behavior reflects the non-Hermitian balance required for edge amplification: stronger \(G\) more efficiently routes atomic dissipation into the cavity mode, overcoming photon leakage and concentrating the local topological density along the edge trajectory.

In Fig.~\ref{SFig3}(b), varying the SOC strength \(\alpha\) from \(1.5\,\Omega\) to \(3.5\,\Omega\) primarily reshapes the \emph{geometry} of the hybrid bands and, with it, the distribution of topological weight. A larger \(\alpha\) enhances spin–momentum locking (\(\propto \alpha k_{x}\sigma_{y}\)), widens the SOC-induced separation of the polaritonic branches, and accentuates their \(k\)-asymmetry. The Chern marker, reconstructed from the same transmission data, responds by developing stronger, more localized peaks that track the gap-crossing edge path in \((k,\omega)\); their positions shift consistently with the SOC-driven displacement of the avoided crossing. 

Together, the two panels establish a clear division of roles: \(G\) controls the \emph{strength and activation} of the topological edge response through cooperativity, while \(\alpha\) controls \emph{where} in \((k,\omega)\) the response concentrates, by setting the band splitting and spectral asymmetry.